\documentclass{aa}

\usepackage{booktabs}  
\usepackage{amsmath}
\usepackage{textcomp}  
\usepackage{xcolor}
\usepackage{gensymb}
\usepackage{lipsum}    
\usepackage{longtable}
\usepackage{natbib}    
\usepackage{orcidlink}

\bibpunct{(}{)}{;}{a}{}{,} 
 
\begin{document}

\title{Time-incremented Multiscale Evolution (TIME): A Code-independent Method for Time-domain 3D Hydrodynamics and its Application to Roche Lobe Overflow}
\titlerunning{Time-incremented Multiscale Evolution (TIME)}

\author{David Dickson\orcidlink{0009-0006-4860-9212}\inst{1,2}}

\offprints{D. Dickson, \email{davey.dickson@kuleuven.be}}

\institute{Institute of Astronomy, KU Leuven,
Celestijnenlaan 200D, 3001, Leuven, Belgium
 \and Department of Physics, North Carolina State University, Raleigh, NC 27695-8202, USA}

\abstract {Many critical physical processes, such as Roche lobe overflow, strain modern simulation methods due to their durations and multidimensionality.} {We aim to create the first 3D model of Roche lobe overflow across evolutionary timescales and use it to characterize the evolution of mass transfer efficiency and timescale.} {Using a novel piecewise approach which alternates between high-resolution 3D dynamic modeling with VH-1 and computationally fast evolutionary modeling, we present and test a method capable of self-scaling variable time resolution at greatly reduced computational cost.} {We find mass transfer in the test high mass x-ray binary M33 X-7 to be unstable and fully conservative in both mass and angular momentum transport onto the accretion disk beyond $f\gtrsim1.01$. This phase begins on thermal timescales and accelerates to span $<100\text{ yrs}$ beyond $f\geq1.1$, while the non-conservative stable phase of $f\lesssim1.01$ occurs on roughly nuclear timescales.} {We identify a critical point $f\sim1.01$ which terminates stable overflow, which may correspond to the point $\dot{M}_{\text{L1}}\sim\dot{M}_{\text{wind}}$ or $\dot{M}_{\text{L1}}\sim10^{-6}M_\odot/\text{yr}$ in the general case.}

\maketitle
\nolinenumbers 


\section{Introduction}
Across many disciplines, 3D hydrodynamic approaches have been coupled with 1D models either as boundary conditions, called geometric multiscale, or initial conditions, called temporal multiscale \citep{mccune2000mixed,formaggia2001coupling,twigt2009ocean,galindo2011auto,weinan2011principles,hemoreview2016,bio2017multitime,joyce2019mesato3d,nordbotten2019review}. Multiscale methods enable the accurate modeling of systems too large or prolonged to be suitable for a fully 3D approach \citep{multiscalereview}. Some temporal multiscale methods extend beyond initial conditions by decoupling system behaviors which operate on disparate timescales \citep{tuckerman1992reversible,flack2012multiple,timescalesbook}. 

A gap remains for systems which require explicit feedback between operations on distinct timescales, such as Roche lobe overflow (RLOF). We propose a novel method of temporal multiscale modeling integrated into a piecewise evolution framework.

\subsection{Application to Roche Lobe Overflow}

We apply this method to the evolution of a donor star undergoing RLOF, the process by which a stellar envelope exceeds its gravitational bounds and falls onto an accretor companion \citep{lubow&shu,paczynski1976close}. The dynamics of RLOF are critical to a range of binary systems, such as gravitational wave sources, common envelopes, and x-ray sources \citep{savonije1978roche,ivanova2013,marchant}. Historically, RLOF models have been limited to either dynamical or time-domain analysis without self-consistent representation of how these feedback, as in e.g. \cite{lubow&shu,bisikalo2000,nazarenko2006,marchant}. Smoothed particle hydrodynamics methods \citep[e.g.][]{passy2011simulating,de2014,iaconi2016effect,reichardt2016strength,reichardt2019extending} have extended multidimensional simulations of RLOF and related close binary interactions across spans of $\sim10\text{ yrs}$, still well below the timescale of stable mass transfer (MT) \citep{marchant2024rev}. 

The stability of MT during RLOF depends critically on the feedback between dynamical and evolutionary processes \citep{stabilitycriteria}. Unstable MT is defined by a runaway feedback loop that rapidly strips the donor star envelope \citep{accretionpower}. Stable MT, by contrast, exhibits gradual mass loss through a tidal stream that self-regulates across evolutionary timescales.

We simulate a test high-mass x-ray binary (HMXB) across the full duration of RLOF. This simulation combines a high-resolution 3D model in VH-1 with a 0D piecewise linear extension, the approach we term as the 3+0D TIME method in Section \ref{ssec:3+0d} \citep{vh1}. We present a series of models of RLOF to characterize its duration and changes with respect to overflow.

\section{Time-incremented Multiscale Evolution} 

A time-incremented multiscale evolution (TIME) model begins by achieving a steady state in a 3D dynamical model and deriving its corresponding evolutionary time increment $\Delta t_{\text{evol}}$ (Section \ref{sec:time}). Then, the system state is mapped to an evolutionary model of lower dimensionality and evolved for a duration $\Delta t_{\text{evol}}$. Closing the loop, an updated 3D dynamical model is initialized from the evolved system. This method may be repeated across the complete duration of the relevant evolution, thereby enabling high-resolution 3D modeling of prolonged system processes.

We assume system evolution to be smooth and continuous, and that both models follow explicit time-stepping schemes.\footnote[3]{This approach may be modified to accommodate implicit time evolution schemes, though that is beyond the scale of this work.} The fractional error this method induces can be minimized with an appropriate time resolution.

\subsection{Variable Time Resolution} \label{sec:time}

Accurately modeling time-domain evolution requires a complete set of characteristic timescales, $\tau_i$, defined with respect to key system parameters, $x_i$ \citep{timescalesbook}. In each 3D step, the model is relaxed to reach a quasi-steady state, periodic fluctuation, or consistent evolution.\footnote[4]{Systems that do not converge may be evolving on too short of timescales for the TIME method. Full 3D may be preferred.} From the mean post-relaxation rate $\langle\dot{x_i}\rangle$, we may define a time step

\begin{equation}
    \label{eqn:generaltimescale}
    \Delta t_i=C_i\tau_i \quad\text{where} \quad\tau_i = \frac{x_i}{|\langle\dot{x_i}\rangle|} \quad .  
\end{equation}

Time steps are scaled by a constant scaling factor, $C_i$, according to the required resolution of the problem.\footnote[5]{Consider the sensitivity of the modeled system to changes in $x_i$ when setting $C_i$.} For systems with additional well-established timescales, these should be computed as well. 

The evolutionary time step $\Delta t_{\text{evol}}$ is then simply equal to the briefest time step $\Delta t_i$ that is longer than the post-relaxation duration of the 3D model. The effects of more rapid timescales are suitably fitted within the span of the 3D model itself.

By recalculating $\tau_i$ for each 3D step, we achieve a variable time resolution $\Delta t_{\text{evol}}$ at which to evolve the system in lower dimensions.

\subsection{3+1D Approach}

In systems for which a 1D model can be meaningfully used for initialization and evolution, the 3+1D TIME method is preferred. This approach utilizes a 1D code for system evolution between successive 3D models. 

This maximizes fidelity in fitting complex evolutionary processes. Additionally, this method offers the advantage of introducing physics relevant to evolution that are not directly relevant to the hydrodynamics. For example, stellar chemical enrichment is generally neglected in dynamical simulations but is essential to stellar evolution.

\subsection{3+0D Approach} \label{ssec:3+0d}

In some cases, it may be necessary to implement a simpler method of piecewise time evolution. Effective 1D models may not exist for a particular system, or may not be compatible with full 3D approaches. 

In 3+0D TIME series, the steady-state variables of the dynamical model are not passed to another simulation. They are instead fed into a piecewise-linear expression that takes the rates of system dynamics to be invariant across a suitably brief $\Delta t_{\text{evol}}$.\footnote[6]{We note this is a conceptual extension of Courant-limited time evolution across a second order of time steps.} 

Even when a 3+1D approach may be a priori preferred, 3+0D exploration may be more effective for preliminary study of previously unexamined regimes.

\section{TIME Modeling of Roche Lobe Overflow}

MT during RLOF depends on high-resolution multidimensional dynamics across the entire binary system. This MT can span timescales $10^6$ times longer than complete 3D approaches, depending on stability \citep{marchant2024rev,stabilitycriteria}. RLOF therefore presents an ideal application for TIME modeling. 

We elect to model RLOF in the simplest case, using a 3+0D approach in absence of evolutionary response from the donor star. This solution is applied to the test HMXB M33 X-7, chosen for its recent observations (Table \ref{tab:m33}). These observations differed from previous findings in MT stability and overfilling factor $f$ \citep{ramachandran2022}.\footnote[7]{The overfilling factor is defined in Appendix \ref{app:roche}.} These uncertainties make M33 X-7 a prime candidate for more detailed study.

Our TIME series begins with a 3D hydrodynamic simulation following the system parameters of M33 X-7 given in Table \ref{tab:m33}.

\begin{table}
\caption{M33 X-7 System Parameters \label{tab:m33}}
\begin{tabular}{ccc}
\hline
Parameter  & Symbol & Value\\
\hline
Donor mass & $M_d$ & $38$ $M_\odot$ \\ 
Donor radius& $R_d$ & $20.5$ $R_\odot$\\
Donor temperature & $T_{\text{eff}}$ & $31$ kK \\
Wind terminal velocity& $V_\infty$ & $1500$ km/s \\
Wind mass loss rate & $\dot{M}_{\text{wind}}$ & $6.3\times 10^{-7}$ $M_\odot/\text{ yr}$ \\
Accretor Mass & $M_a$ & $11.4$ $M_\odot$ \\
Orbital period & $P_{\text{orb}}$ & 3.45 days \\
Binary separation & $d$ &  $35$ $R_\odot$\\
\hline
\multicolumn{3}{c}{Values taken from \cite{ramachandran2022}.}
\end{tabular} 
\end{table}

\subsection{Dynamical Simulation Methodology}
\label{ssec:methods}

We employ the 3D hydrodynamics code VH-1, which solves Euler's equations for an ideal gas by the piecewise parabolic method \citep{ppm,vh1}. VH-1 applies angular momentum (AM) conservation, radiative cooling, and a line-driven wind across a specialized set of overlaid spherical yin-yang grids \citep{sobolev,yinyang,dickson2024thesis}. We employ the same method as \cite{dickson2024}, except for modifications to the disk approach, grid, and thermalization.

As in \cite{dickson2024}, we fit a 1D stellar profile by potential and map it onto the Roche geometry of the donor. This approach assumes a corotating donor near hydrostatic equilibrium\footnote[8]{Exception being for the region near the L1 point, which cannot achieve hydrostatic equilibrium.}. Outside the donor, we initialize only donor wind, forming an accretion disk during model relaxation.

We set our accretor inner boundary at $0.039d\sim1 R_\odot$, within the natural band gap identified in \cite{dickson2024}. We also employ a reduced thermalization timescale of $\Delta t_{\text{settling}} = 1$ second for computational efficiency. In preliminary trials, we found the steady state invariant to these changes.

\subsection{Time Integration Methodology} \label{ssec:int}

We begin our TIME series from RLOF onset (an overfilling factor of $f=1.0005$) to examine the system across a range of overflows. We call this model A, denoting subsequent dynamical steps in alphabetical order.

Each individual 3D model is evolved for a simulated $\sim1 \text{ day}$ after achieving a quasi-steady state. We then extract mean post-relaxation mass and AM fluxes across spherical surfaces about the donor, accretor, and binary. These yield the Roche timescale $\tau_{\text{Roche}}$ as given by 

\begin{equation}
    \tau_{\text{Roche}} = \left| \frac{\dot{M}_{\text{tot}}}{M_{\text{tot}}} + 2\frac{\dot{J}}{J} - 2\frac{\dot{M}_d}{M_d} - 2\frac{\dot{M}_a}{M_a} + Q\left(\frac{\dot{M}_a}{M_a} - \frac{\dot{M}_d}{M_d}\right) \right|^{-1} \text{\quad.}
    \label{eqn:roche}
\end{equation}

where $Q$ is a function of the mass ratio.\footnote[9]{A derivation of $\tau_{\text{Roche}}$ and $Q$ is provided in Appendix \ref{app:roche}.}

VH-1 is artificially slow to dissipate AM within the disk, as it lacks magnetism and viscosity. We therefore defer to the net inflow of mass and AM through the spherical surface surrounding the disk region to define the accretion rate.

After each 3D step, the system is evolved by $\Delta t_{\text{evol}}=0.01\times\tau_{\text{Roche}}$ such that the overfilling factor varies by $\sim1\%$ between successive models; this high resolution is critical as prior research suggests the system to be highly responsive to small changes in overflow \citep{dickson2024}. 

The masses and total angular momentum are updated linearly by the mean fluxes obtained. A new binary separation is then computed by Equation \ref{eqn:qANDd}. We hold constant the eclipse radius of the donor and numerically calculate the new overfilling factor for that radius. Finally, the subsequent 3D model is initialized as before, with the exception of the disk.

While disk formation occurred within the simulation in our early models, later models could not circularize within simulation time. We address this with a change of method beginning with model J. Thereafter, each model initializes the accretor grids by the end state of the previous model.

Lacking the donor response, we here define MT as stable when it is slow enough to be comparable to the donor nuclear timescale.

\subsection{Efficiencies} \label{ssec:eff}

Across this evolution, we track four efficiencies of MT, as previously employed in \cite{dickson2024}. These are respectively the mass and AM transfer efficiencies of the binary as a whole,

\begin{equation}
    \label{eqn:efftot}
    \alpha_\text{MT(tot)} = \frac{\dot{M}_{a}}{-\dot{M}_{d}} \quad\text{and}\quad \alpha_\text{AM(tot)} = \frac{\dot{L}_{a}}{-\dot{L}_{d}} \quad ,
\end{equation}

and those of the L1 tidal stream in particular,

\begin{equation}
    \label{eqn:effL1}
    \alpha_\text{MT(L1)} = 1-\frac{\dot{M}_{\text{s,out}}}{\dot{M}_{\text{L1}}} \quad\text{and}\quad \alpha_\text{AM(L1)} = 1 -\frac{\dot{L}_{\text{s,out}}}{-\dot{L}_{\text{L1}}} \quad .
\end{equation}

Values denoted by "s,out" refer to the nonconservative outflow stream formed from the collision of the L1 tidal stream with the accretion disk. These efficiencies do not account for transport after entry into the accretion disk.

\section{Results} \label{sec:results}

\begin{figure}
    \centering
    \includegraphics[width=0.95\linewidth]{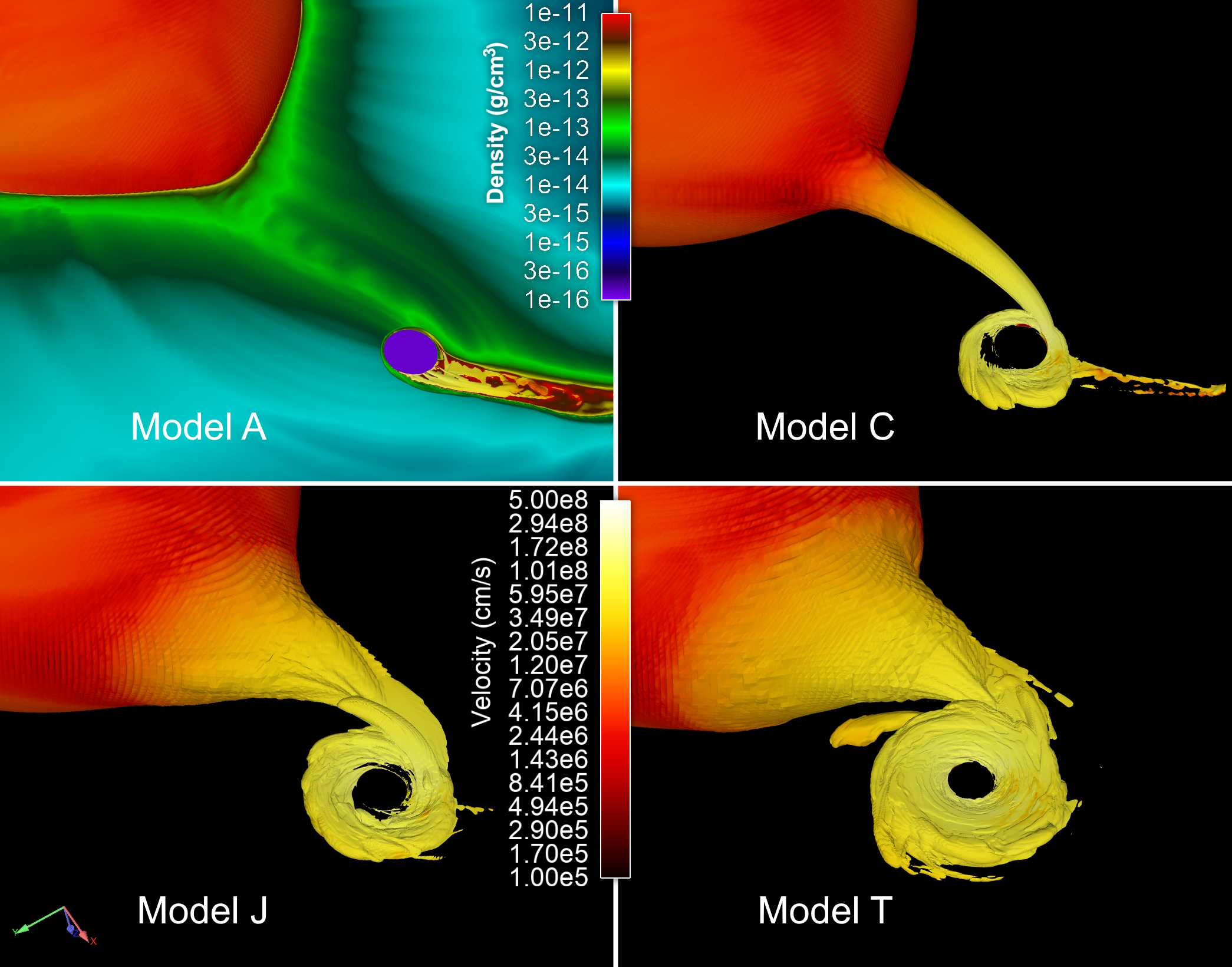}
    \caption[Binary System Visualized by Sequential Simulations]{Binary System Visualized by Sequential Simulations. The donor, tidal stream, and accretion disk are visualized by an isodensity surface for select models. The equatorial plane is additionally mapped for model A, as no tidal stream formed. For the corresponding visualizations of all twenty models, see Appendix \ref{app:figs}.}
    \label{fig:sequentialbig}
\end{figure}

To study the efficiency, rate, and timescale of RLOF across evolution, we performed a 3+0D TIME series spanning from the onset of full RLOF to the final decades of unstable MT.

Our TIME series of twenty models spanned a total of $616,307$ yrs, comparable to the nuclear timescale of the donor \citep{kippenhahnweigert}. During that span, the system evolved piecewise from $f=1.0005$ to $f=1.1779$, nearly halving the volume of the Roche lobe. This constituted $<3\%$ of the Roche timescale of model A, indicating substantial sensitivity to changes in overflow. This series is summarized in Table \ref{tab:serialset} and visualized in Figure \ref{fig:sequentialbig}. 

Each 3D model spanned $\sim1$ day of simulated time in quasi-steady state, meaning this approach incurred $10^7$ times less computational expense than a purely hydrodynamic model of the same $616$ kyrs.

We chose to end our series after twenty dynamical models due to exponentially briefer timescales, grid limitations, and the span of dynamical thresholds covered, most notably exceeding the L2 potential and eclipse inflection.\footnote[10]{The eclipse inflection is the point at which the tidal stream eclipses the donor in the accretor's sightline \citep{dickson2024thesis}}.

Our simulations begin with the unexpected absence of full RLOF in model A.

\begin{table}
    \centering
    \caption{Evolution of Orbital Configuration by Simulation \label{tab:serialset}}
    \begin{tabular}{ccccc}
    \hline
     Model & $f$ & Mass  & Separation  &  Time  \\
     & & Ratio& ($R_\odot$) &  (Years)\\
    \hline
A & 1.0005 & 3.33 & 35.3 & 0 \\
B & 1.0083 & 3.31 & 35.0 & 279,432 \\
C & 1.0163 & 3.29 & 34.8 & 545,157 \\
D & 1.0246 & 3.26 & 34.6 & 601,652 \\
E & 1.0329 & 3.24 & 34.3 & 611,135 \\
F & 1.0416 & 3.22 & 34.1 & 614,111 \\
G & 1.0501 & 3.20 & 33.9 & 615,195 \\
H & 1.0587 & 3.18 & 33.6 & 615,682 \\
I & 1.0673 & 3.16 & 33.4 & 615,931 \\
J & 1.0762 & 3.14 & 33.2 & 616,066 \\
K & 1.0849 & 3.12 & 33.0 & 616,145 \\
L & 1.0941 & 3.10 & 32.8 & 616,196 \\
M & 1.1030 & 3.08 & 32.5 & 616,229 \\
N & 1.1120 & 3.06 & 32.3 & 616,252 \\
O & 1.1210 & 3.04 & 32.1 & 616,268 \\
P & 1.1303 & 3.02 & 31.9 & 616,279 \\
Q & 1.1394 & 3.00 & 31.7 & 616,288 \\
R & 1.1487 & 2.98 & 31.5 & 616,294 \\
S & 1.1581 & 2.96 & 31.3 & 616,299 \\
T & 1.1681 & 2.94 & 31.0 & 616,303 \\
End State & 1.1779 & 2.92 & 30.8 & 616,307 \\
\hline
    \end{tabular}
\end{table}

\subsection{Wind Roche Lobe Overflow at $f\gtrsim1$}

Model A does not launch a dense tidal stream as its overfilling factor would indicate. We find model A remains in the wind RLOF phase with an enhanced wind transferred through the L1 region onto the BH accretor (Figure \ref{fig:sequentialbig}). The accretor of Model A captures $3\%$ of the donor wind, $17\times$ the rate of Bondi-Hoyle-Lyttelton wind accretion \citep{EDGAR2004bhl}. We do not simulate close enough to the BH to resolve a bow shock as is generally expected and present in similar models \citep{blondin2012hoyle,blondin2024mass}. We therefore take our accretion rate as an upper limit well within the expected range of wind RLOF \citep{mohamed2007wind}. 

\begin{figure}
    \centering
    \includegraphics[width=0.75\linewidth]{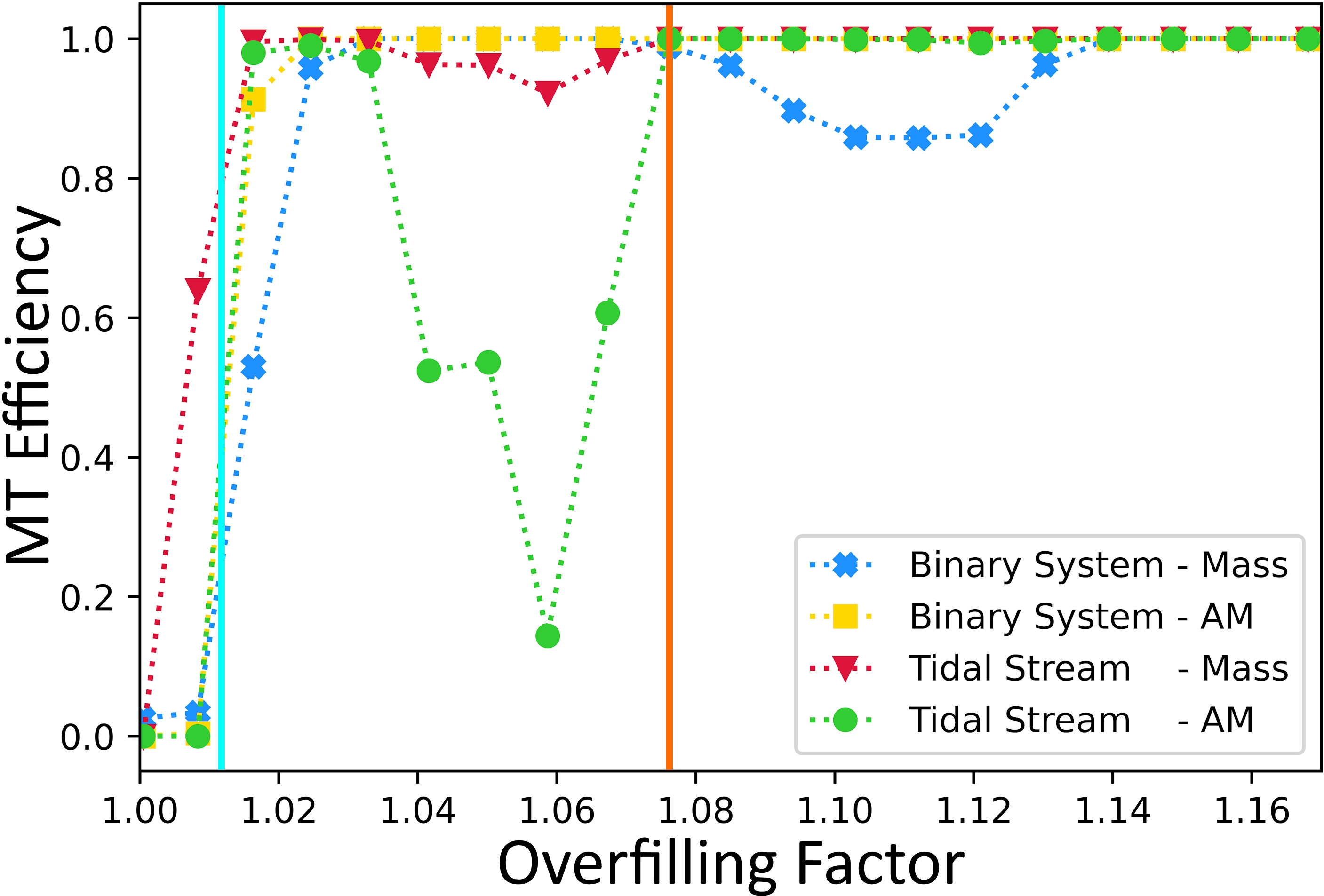}
    \caption[Mass Transfer Efficiency vs. Overfilling Factor]{Mass Transfer Efficiency vs. Overfilling Factor. Depicted are the efficiencies of mass and angular momentum (AM) transport across system evolution defined in Section \ref{ssec:eff}. The cyan bar indicates the instability threshold $f \sim 1.01$. The orange bar indicates the modification made to disk retention with model J noted in Section \ref{ssec:int}.}
    \label{fig:efficiency}
\end{figure}

In the absence of a bow shock, a dense stream of collapsed wind outflowed from the wake of the accretor. This accretor outflow stream carried away nearly all of the AM of the accreted wind (Figure \ref{fig:efficiency}).

\cite{el2019wind} modeled a similar HMXB with a BH accretor at a mass ratio of 2 and filling factor $f=0.9$. They estimated the BH to capture $7\%$ of the donor wind at our wind velocities, $\sim10\times$ higher than predicted by wind accretion alone. While their system differs on several key parameters, the wind capture rate they obtained is comparable to our result of $3\%$ and a $17\times$ wind enhancement. This suggests wind RLOF may occur even in the $f\gtrsim 1$ regime, rather than strictly defined by the limit $f<1$ as previously suggested \citep{mohamed2007wind,el2019wind}.

\cite{el2019wind} suggest wind RLOF provides a consistently stable alternative to RLOF-induced MT in high mass ratio binaries. They define the stable regime by $\dot M_d \leq 10^{-6} M_\odot/\text{yr}$, which includes our model A ($\dot M_{d,A} = 8.5 \times 10^{-7} M_\odot/\text{yr}$). The Roche timescale of model A was $27.9 \text{ Myrs}$, exceeding the donor's nuclear timescale. 

Despite both of these stability regimes being satisfied, our wind RLOF model led directly to the onset of unstable MT. In absence of a mediating stellar response, our method makes eventual unstable MT inevitable. However, the sub-nuclear timescale in which instability formed suggests wind RLOF may induce unstable MT, as proposed by \cite{sun2023effects}. They found low mass binaries of similar mass ratios to M33 X-7 more often resulted in unstable MT when they accounted for wind RLOF. Our results validate this discrepancy.

\subsection{Stable Mass Transfer}

In model B, stable MT begins with the formation of a tidal stream. At L1, this tidal stream exceeds the density at which the wind launches \citep{dickson2024}. Beyond L1, the tidal stream becomes diffuse, comparable to a wind overdensity. Physical systems similar to model B may therefore bear observational resemblances to wind RLOF systems. Even diffuse, this tidal stream was moderately efficient in mass accretion (Figure \ref{fig:efficiency}). However, model B still fails to deposit significant AM onto the accretor. 

While the stable MT of model B was nonconservative, this changes rapidly with model C. Its tidal stream is dense throughout and carries $92\%$ as much mass as the donor wind. Throughout the remainder of the stable MT regime, mass and AM transfer through the tidal stream are nearly conservative.

Disk formation in model C raised AM efficiency from $<1\%$ to $91\%$. This indicates tidally-fed disks are an order of magnitude more effective at capturing AM than wind accretion or wind RLOF per unit mass.

Models A and B exhibited almost identical Roche timescales on the order of the nuclear timescale independent of L1 tidal stream formation.

\subsection{The Instability Threshold}
\label{ssec:threshold}

\begin{figure}
    \centering
    \includegraphics[width=1.0\linewidth]{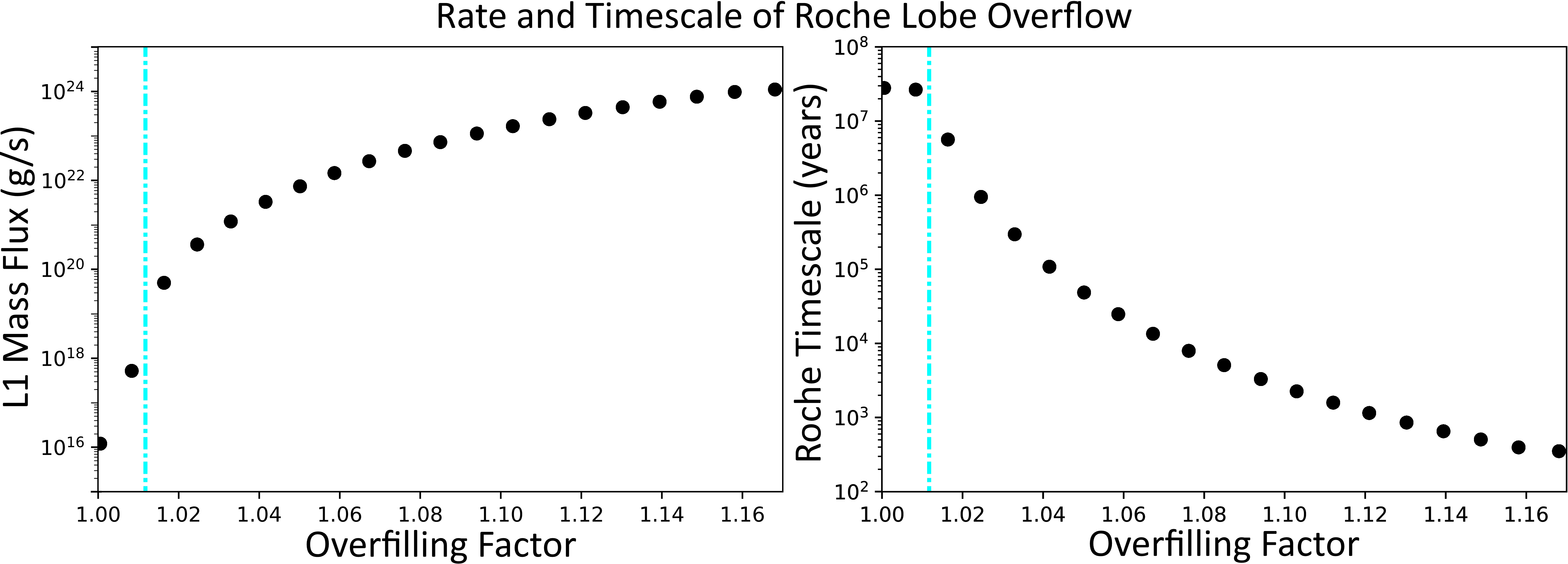}
    \caption[L1 Mass Loss Rate and Roche Timescale vs. Overfilling Factor]{L1 Mass Loss Rate and Roche Timescale vs. Overfilling Factor. The cyan line marks the instability threshold $f \sim 1.01$ detailed in Section \ref{ssec:threshold}. }
    \label{fig:rlo}
\end{figure}

\begin{figure}
    \centering
    \includegraphics[width=1.0\linewidth]{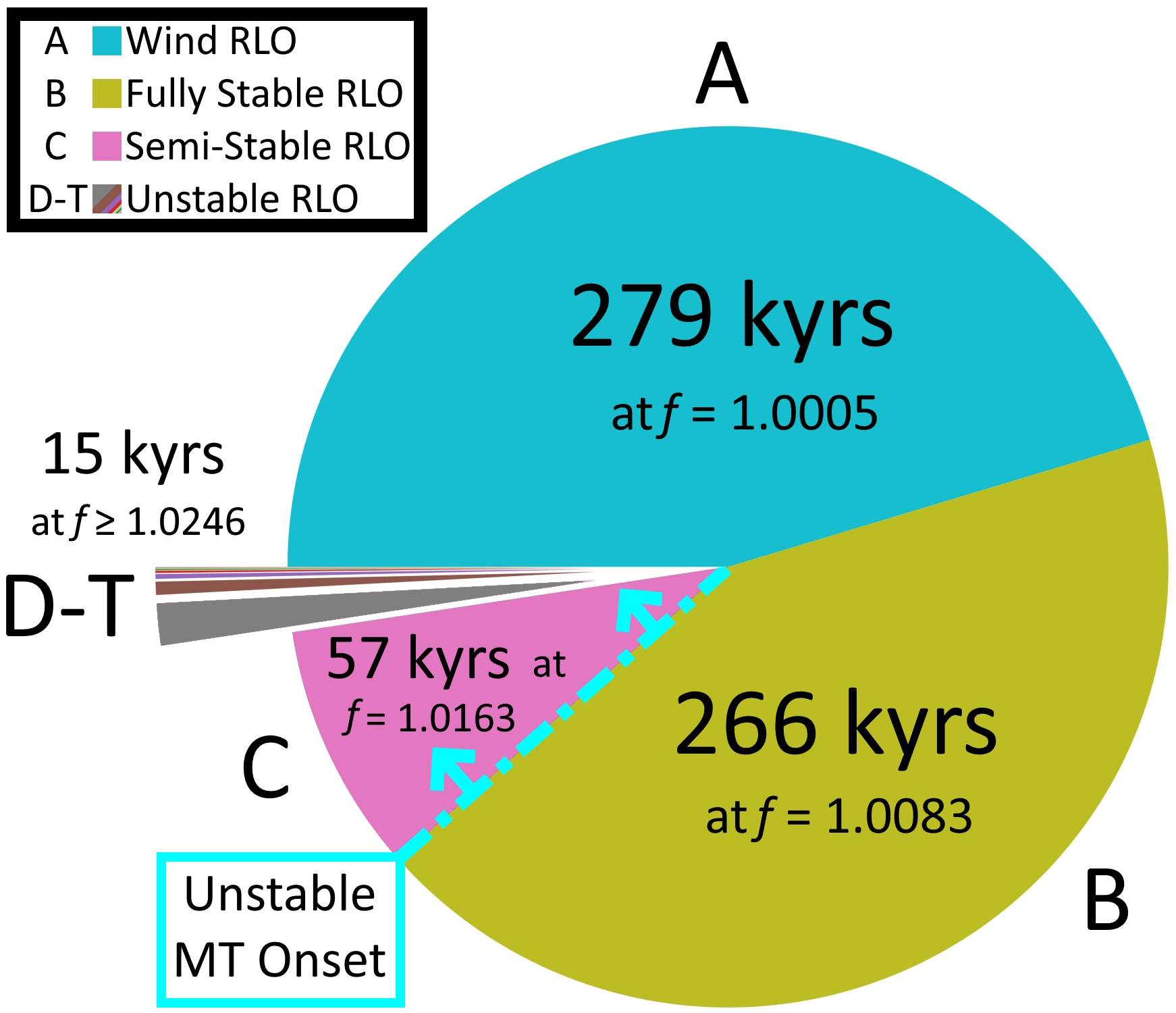}
    \caption[Time Distribution of Sequential Simulations]{Time Distribution of Sequential Simulations. We color models by their RLOF regime (top left) and group models D through T due to their brief durations. The cyan line marks the instability threshold $f \sim 1.01$ detailed in Section \ref{ssec:threshold}.}
    \label{fig:pie}
\end{figure}

We see a distinct shift in system evolution between models B and C, marked by a cyan bar in Figures \ref{fig:efficiency}, \ref{fig:rlo}, and \ref{fig:pie}. Roche timescales accelerated precipitously for all models beyond this threshold (Figures \ref{fig:rlo} and \ref{fig:pie}). This is paired with a rapid rise in mass flux through L1 (Figure \ref{fig:rlo}). 

We therefore take this threshold to define the onset of increasingly unstable MT. This transition occurs near $f\sim1.01$ in our model and coincides with a number of possible system thresholds:

\begin{itemize}
    \item The Roche timescale becomes dominated by the tidal stream, reasonably approximated by conservative transfer that neglects wind.\footnote[11]{This approximate timescale is derived in Appendix \ref{app:roche}.}
    \item MT becomes approximately conservative.
    \item The tidal stream exceeds both the mass loss rate of the wind and the conventional threshold $10^{-6} M_\odot/\text{yr}$ of e.g. \cite{mohamed2007wind}.
    \item A disk forms on-grid. 
\end{itemize}

It is unclear which of these, if any, trigger the instability threshold and which result from it. Future research is required to parameterize the instability threshold.

The first three models after the instability threshold (C-E) are arguably stable, spanning $\sim70$ kyrs with Roche timescales within an order of the nuclear timescale (Figure \ref{fig:pie}). The remaining 15 models (F-T) exhibit distinctly unstable MT across a span of only $\sim2$ kyrs. This feedback loop propels the system into ever-increasing overflow and L1 mass fluxes that exceed $10^{-2} M_\odot/\text{yr}$ by model T. 

\subsection{Unstable Mass Transfer at High Overflow}
\label{ssec:unstable}

We found unstable MT to be conservative in both mass and AM transfer onto the accretor region, consistent with the $f=1.1$ model of \cite{dickson2024}. Figure \ref{fig:efficiency} reflects this conclusion, save for two spurious features. 

In models D-I (left), tidal stream efficiency is artificially suppressed due to incomplete disk circularization.\footnote[12]{This was counteracted in later models with a modification of technique described in Section \ref{ssec:methods}} Conversely, models J-T (right) form wide disks beyond the accretor-centered flux surface. These effects lead to apparent deficits that do not represent escaping mass. We therefore conclude unstable MT to be fully conservative in both mass and AM across our TIME series. 

The MT rate of the unstable regime is qualitatively in keeping with the results of \cite{savonije1978roche} for low mass binaries of comparable mass ratios and periods. With a simple prescription for RLOF, they found similar curvature to Figure \ref{fig:rlo} in the evolution of the MT rate with respect to time. This may change beyond $f=1.17$, though model T suggests otherwise.

The donor surface of model T surpasses not only the eclipse inflection but the equipotential of L2 \citep{dickson2024thesis}. This theoretically enables initially hydrostatic donor material to escape the accretor lobe at much higher rates. Model T shows no significant variation from the trends of previous models despite its extreme overflow.

\subsection{Span of Simulations}
\label{ssec:span}
At its current trajectory, the simulated system cannot sustain RLOF for more than $15$ yrs beyond our final model (T). Our TIME series therefore models $\geq99.997\%$ of the RLOF phase from onset to the final years prior to probable common envelope formation. An additional $\sim 100$ dynamical simulations would be required to span the remaining years, due to exponentially faster system feedback (Figures \ref{fig:rlo} and \ref{fig:pie}).

Our model suggests M33 X-7 cannot sustain overflow of $f\geq 1.1$ for more than 100 yrs. The observational window for high overflow, like that predicted by \cite{ramachandran2022}, is therefore very brief. A secondary tidal stream at the outer Lagrange point of the donor, as suggested by \cite{marchant}, is here predicted to last an even briefer $<10\text{ yrs}$ in this configuration.

Our method overestimates timescale and underestimates mass loss in the final $\sim30\text{ yrs}$ of our TIME series. This stems from our assumptions of synchronous corotation and thermal equilibrium, which break down in the rapid evolution of high overflow \citep{packet1981spin,hjellming1987thresholds}. The variations in timescale these may produce are therefore secondary to the limitations of our time resolution.

We probed a brief second TIME series with $\Delta t_{\text{evol}}=0.1\%\ \tau_{\text{Roche}}$. This $10\times$ increased time resolution reduced the estimated timescale by as much as $\sim42\%$ about the instability threshold, where the piecewise-linear approximation deviates most. We therefore assert the Roche timescales we present for the unstable MT regime as upper limits within an order of magnitude of the true values for this configuration.

\section{Conclusions}

We present a novel and generic method of time-domain multidimensional dynamical simulations. This method is vastly more computationally accessible than pure hydrodynamics by decoupling the total time spanned by a TIME series from the Courant-Friedrichs-Lewy condition. The TIME method therefore allows arbitrarily high resolution to be employed across arbitrarily long physical processes in 3D. 

We demonstrate the use of this methodology to evolve an overflowing binary in 3D for the first time without prescriptive MT. Our simulation simultaneously models the donor envelope, photosphere, wind, and L1 tidal stream, as well as the accretion disk surrounding the BH accretor, on a non-uniform grid. 

We implement a TIME series of twenty models to examine system configuration, fluxes, timescales, and efficiencies across an overfilling factor range $1.0005 \leq f \leq 1.1681$. This was achieved by the 3+0D TIME method with adaptive time steps of $0.01 \times \tau_{\text{Roche}}$ (Equation \ref{eqn:roche}).

We vary overfilling factor across nearly the entirety of a simulated MT phase absent donor evolutionary response, spanning from wind RLOF prior to full RLOF onset to within $15$ yrs of common envelope formation.

\begin{enumerate}
    \item We utilize the TIME method to examine RLOF across previously computationally inaccessible regimes. Compared to a full 3D approach, we reduced computational cost by a factor $10^7$ to $10^{4}$ core-hours across its $616$ kyr span.
    \item We provide the first model of MT efficiencies as they vary with overfilling factor. We find unstable MT to be fully conservative in both mass and AM transport onto the accretion disk. We find stable MT to become conservative (in mass) as it approaches $\dot M_{\text{L1}} \gtrsim \dot M_{\text{wind}}$. Stable MT remained inefficient in transporting AM, as shown in Figure \ref{fig:efficiency}. This calculation excludes the efficiency of transport after deposition onto the disk.
    \item We identify an instability threshold at $f\sim 1.01$ and $\dot M \sim 10^{-6}M_\odot/\text{yr}$ in our system, beyond which MT becomes increasingly unstable. This coincides with multiple system thresholds, as detailed in Section \ref{ssec:threshold}.
    \item We provide a preliminary model of the timescale of RLOF, when considered in absence of donor radial response. Both wind RLOF and stable full RLOF occurred on nearly nuclear timescales. Unstable MT began on thermal timescales and became exponentially faster at high overflow. The $f\geq1.1$ regime is exceptionally brief, spanning no more than $100$ yrs. This renders observations of high-overflow systems and secondary tidal streams unlikely in systems comparable to M33 X-7.
    \item Wind RLOF can dominate binary interaction at low overfilling factors $f>1$. We observe wind RLOF at $f = 1.0005$ launching with a system MT efficiency of $\leq3\%$. While stable, it directly contributed to eventual unstable MT onset. Models which incorporate wind RLOF may be more likely to trigger unstable MT.
    \item M33 X-7 is likely to undergo stable MT on nuclear timescales followed by an unstable MT phase, unless interrupted by donor evolution.
\end{enumerate}

\section*{Future Work}

Future work may apply the TIME method to a wide variety of multidimensional time-domain systems. The model of RLOF presented here could be improved substantially by a 3+1D TIME approach that accounts for the donor response and by incorporation of Eddington x-ray feedback. A range of TIME series across RLOF regimes could also provide a better constraints on the instability threshold and greater perspective on the impact of multidimensional processes on binary evolution.

\section*{Acknowledgements}

We are grateful for the guidance and mentorship of John Blondin and Jon Sundqvist. We acknowledge support from NSF grant AST-2308141 and from KU Leuven C1 grant BRAVE C16/23/009.


\bibliography{Dickson-TIME}
\bibliographystyle{aa}


\begin{appendix}
\section{Additional Figures}
\label{app:figs}
\begin{figure}
    \centering
    \includegraphics[width=1.0\linewidth]{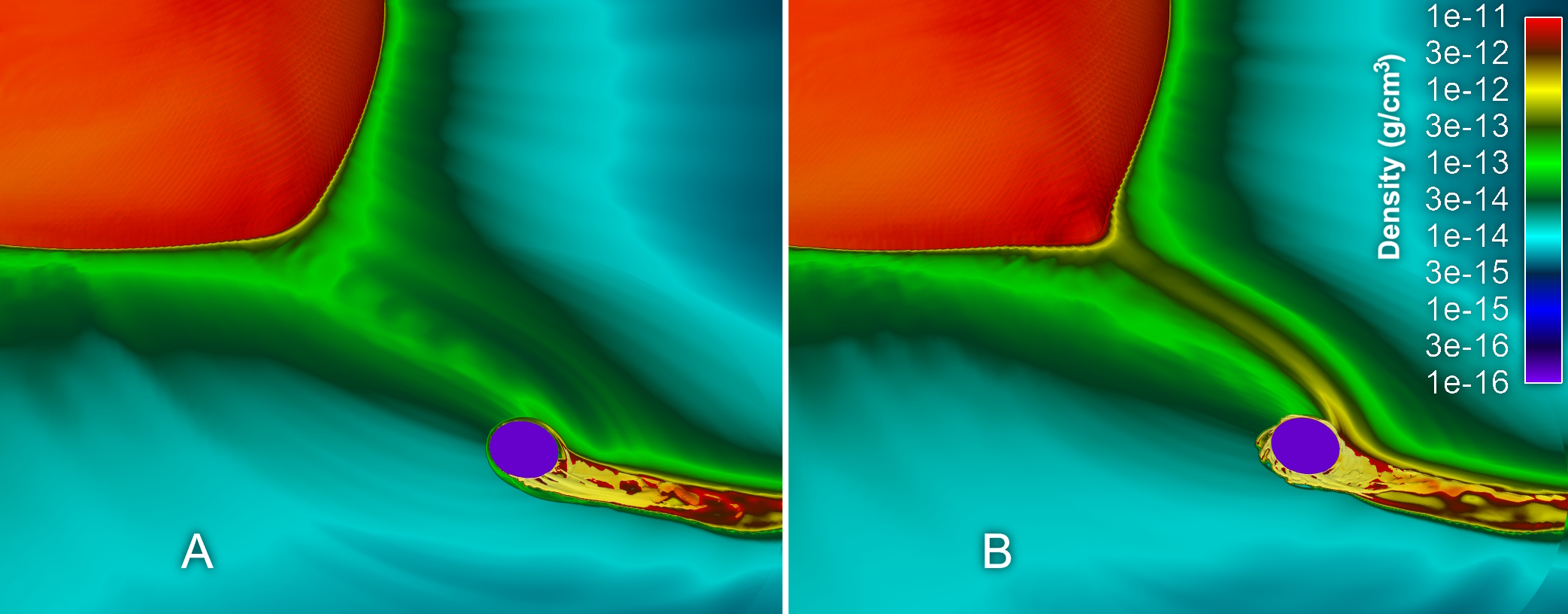}
    \caption[Binary System Visualized for Models A-B]{Binary System Visualized for Models A-B}
    \label{fig:sequentialA}
\end{figure}

\begin{figure}
    \centering
    \includegraphics[width=1.0\linewidth]{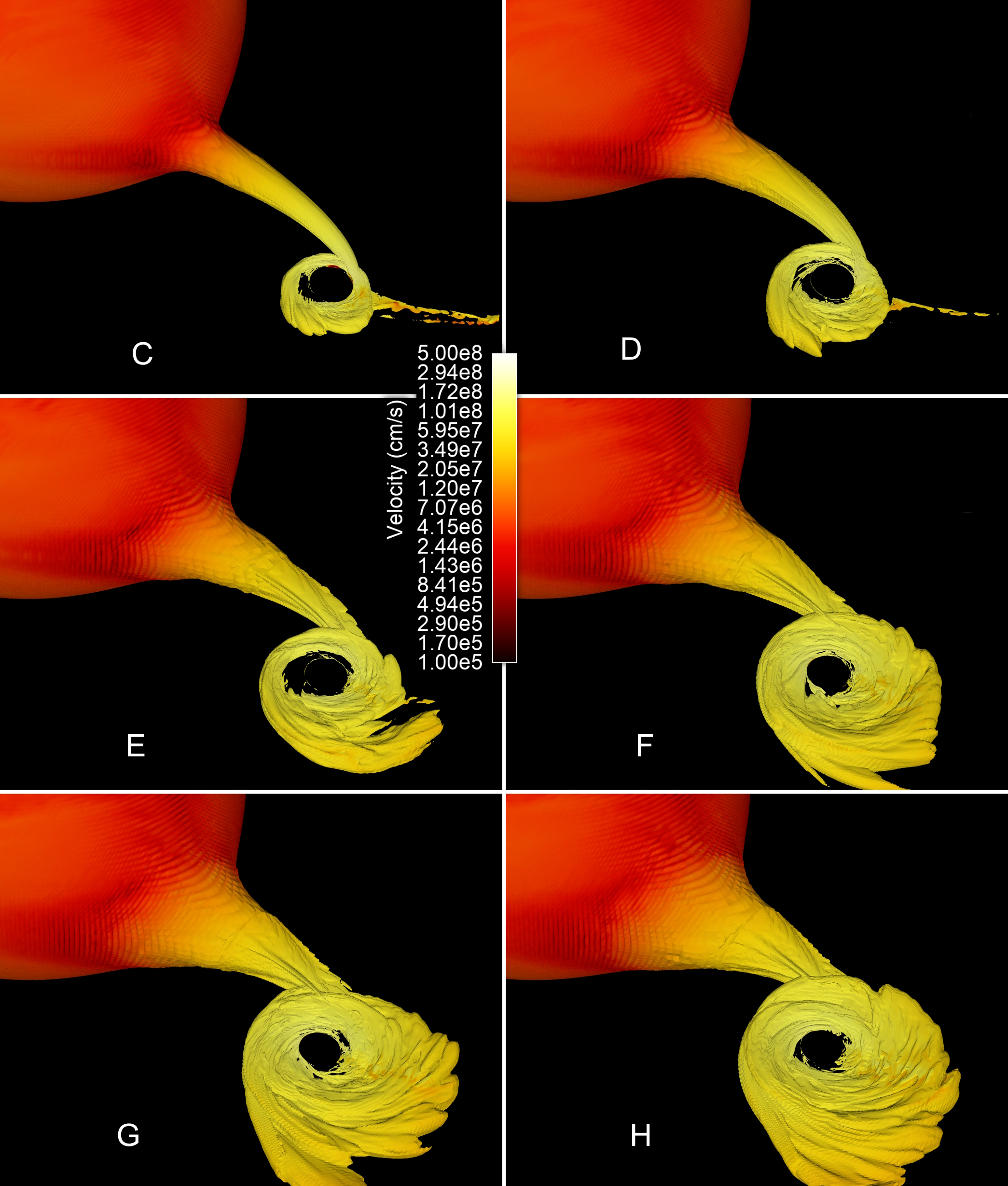}
    \caption[Binary System Visualized for Models C-H]{Binary System Visualized for Models C-H}
    \label{fig:sequentialC}
\end{figure}

\begin{figure}
    \centering
    \includegraphics[width=1.0\linewidth]{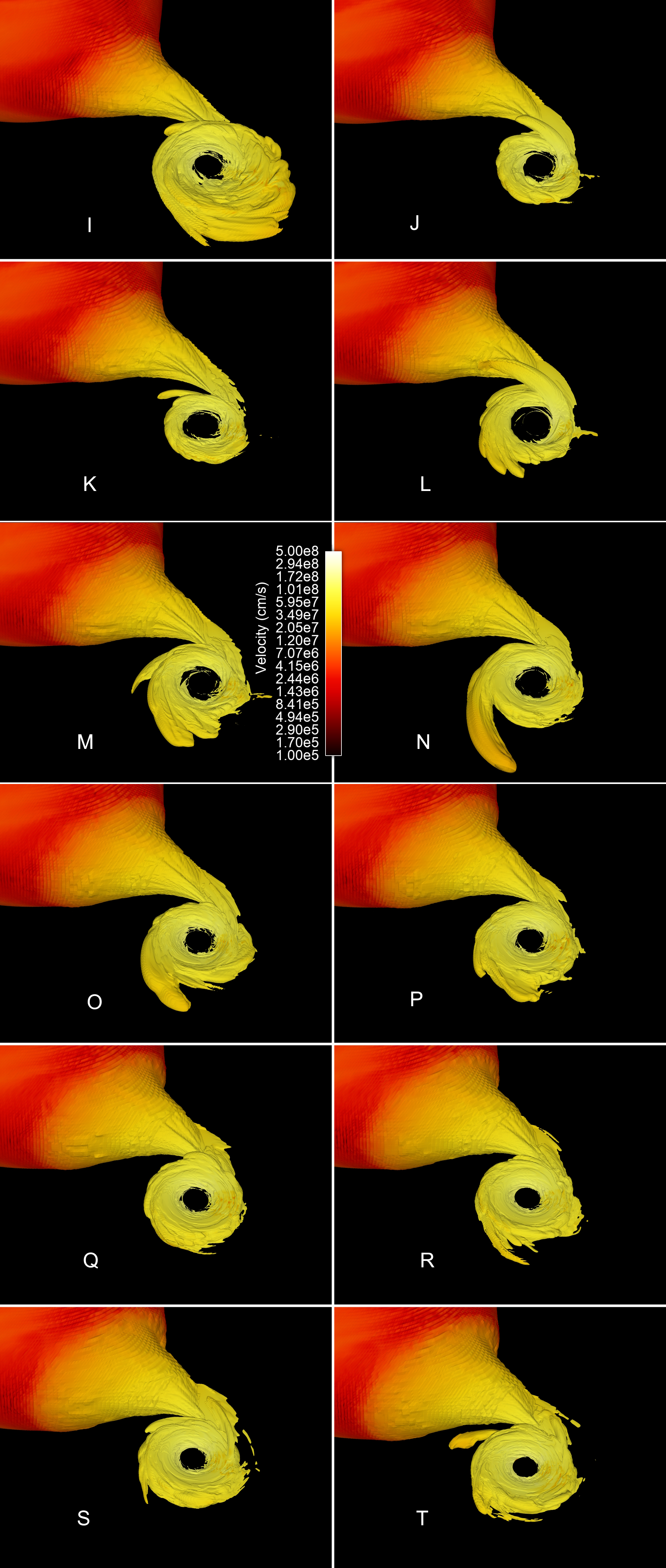}
    \caption[Binary System Visualized for Models I-T]{Binary System Visualized for Models I-T}
    \label{fig:sequentialI}
\end{figure}

We here present all twenty sequential models as fit by the same isodensity surface $(10^{-11}g/cm^3)$ and color gradient also employed in Figure \ref{fig:sequentialbig}. Unlike model A, model B exceeds the density at which the wind launches at L1; however, its stream becomes more diffuse throughout its approach to the BH. This necessitates additional visualization in the ecliptic plane given in Figure \ref{fig:sequentialA}. Figures \ref{fig:sequentialC} and \ref{fig:sequentialI} span the remaining 18 simulations, which do not require the same ecliptic visualization.

\section{Roche Timescale}
\label{app:roche}
We begin with a volumetric Roche lobe radius sufficiently approximated to within $1\%$ by the definition of \cite{eggleton1983} as

\begin{equation}
    \label{eqn:r}
    R_{\text{donor Roche lobe}} \approx d \frac{0.49q^{2/3}}{0.6q^{2/3}+\ln{(1+q^{1/3})}}
\end{equation}

from which we can obtain a first logarithmic derivative with respect to time of

\begin{equation}
    \label{eqn:rdot1ANDqterm}
    \frac{\dot{R}}{R} = \frac{\dot{d}}{d} -Q\frac{\dot{q}}{q} \text{ \quad where \quad} Q = \frac{\frac{1}{1+q^{-1/3}}-2\ln{(1+q^{1/3})}}{1.8q^{2/3}+3\ln{(1+q^{1/3})}}
\end{equation}

in which we have defined $R\equiv R_{\text{donor Roche lobe}}$ and dropped the approximation symbol as within a $1\%$ margin \citep{eggleton1983}. Dotted values refer to mean flux rates rather than instantaneous rates of change. Assuming a Keplerian binary, we may expand this definition using

\begin{equation}
\label{eqn:qANDd}
      J = M_aM_d\sqrt{\frac{Gd(1-e^2)}{M_{\text{tot}}}} \text{ \quad , \quad} q = \frac{M_d}{M_a} = \frac{M_{\text{donor}}}{M_{\text{accretor}}}
\end{equation}

and their logarithmic derivatives with respect to time

\begin{equation}
    \label{eqn:qdotANDddot}
     \frac{\dot{d}}{d} = \frac{\dot{M}_{\text{tot}}}{M_{\text{tot}}} + 2\frac{\dot{J}}{J} - 2\frac{\dot{M}_d}{M_d} - 2\frac{\dot{M}_a}{M_a} + 2\frac{e\dot{e}}{1-e^2} \text{ , }   \frac{\dot{q}}{q} = \frac{\dot{M}_d}{M_d} - \frac{\dot{M}_a}{M_a}
\end{equation}

to obtain the form

\begin{equation}
    \label{eqn:rdot2}
    \frac{\dot{R}}{R} = \frac{\dot{M}_{\text{tot}}}{M_{\text{tot}}} + 2\frac{\dot{J}}{J} - 2\frac{\dot{M}_d}{M_d} - 2\frac{\dot{M}_a}{M_a} + Q\left(\frac{\dot{M}_a}{M_a} - \frac{\dot{M}_d}{M_d}\right) \text{\quad.}
\end{equation}

We here drop the eccentricity term, as we assume a circular corotating binary. We note that this logarithmic derivative provides a characteristic timescale, $\tau_{\text{Roche}}$, of RLOF defined directly by system fluxes 

\begin{equation}
    \label{eqn:tau}
    \tau_{\text{Roche}} = \frac{R}{|\dot{R}|} \quad .
\end{equation}

This enables us to quantify the evolution of the overfilling factor $f$, which is defined by the ratio of the donor star radius and its Roche lobe radius,

\begin{equation}
    \label{eqn:f}
    f = \frac{R_{d}}{R} \quad.
\end{equation}

For the purposes of this work, we take $R_d$ to be constant. This definition of $f$ requires a choice of how to define the radius of an aspherical object. We here measure overfilling factor by the ratio of eclipse radii when taken without the tidal stream present.

If we consider the system to be dominated by conservative MT, as described in Section \ref{ssec:threshold}, we may take

\begin{equation}
    \label{eqn:ml1}
    \dot{J}\approx0\quad\text{,}\quad\dot{M}_{\text{tot}}\approx0 \quad \text{ and} \quad \dot{M}_{\text{L1}} \approx \dot{M}_a \approx -\dot{M}_d \quad .
\end{equation}

Combining this with Equation \ref{eqn:rdot2} yields

\begin{equation}
    \label{eqn:rdot3}
    \frac{\dot{R}}{R} \approx \dot{M}_{\text{L1}}\left(2\frac{M_a-M_d+QM_{\text{tot}}}{M_aM_d} \right) \quad,
\end{equation}

\end{appendix}

\end{document}